\documentclass[11pt,amsmath,superscriptaddress,prl,endfloats,notitlepage]{revtex4-1}
\usepackage{bm}
\usepackage{graphicx}
\usepackage[usenames,dvipsnames,svgnames,table]{xcolor}
\usepackage[colorlinks=true,linkcolor=RoyalBlue,citecolor=RoyalBlue]{hyperref}

\begin{document}
	
\title{Antiferromagnetic Spin Wave Field-Effect Transistor}

\author{Ran Cheng}
\affiliation{Department of Physics, Carnegie Mellon University, 5000 Forbes Avenue, Pittsburgh, PA 15213}
	
\author{Matthew W. Daniels}
\affiliation{Department of Physics, Carnegie Mellon University, 5000 Forbes Avenue, Pittsburgh, PA 15213}

\author{Jian-Gang Zhu}
\affiliation{Department of Electrical and Computer Engineering, Carnegie Mellon University, 5000 Forbes Avenue, Pittsburgh, PA 15213}

\author{Di Xiao}
\affiliation{Department of Physics, Carnegie Mellon University, 5000 Forbes Avenue, Pittsburgh, PA 15213}

 
\begin{abstract}
In a collinear antiferromagnet with easy-axis anisotropy, symmetry dictates that the spin wave modes must be doubly degenerate.  Theses two modes, distinguished by their opposite polarization and available only in antiferromagnets, give rise to a novel degree of freedom to encode and process information.  We show that the spin wave polarization can be manipulated by an electric field induced Dzyaloshinskii-Moriya interaction and magnetic anisotropy.  We propose a prototype spin wave field-effect transistor which realizes a gate-tunable magnonic analog of the Faraday effect, and demonstrate its application in THz signal modulation.  Our findings open up the exciting possibility of digital data processing utilizing antiferromagnetic spin waves and enable the direct projection of optical computing concepts onto the mesoscopic scale.
\end{abstract}

\maketitle
   
Spin waves are propagating spin precessions in magnetically ordered media. Since spin waves can carry pure spin currents in the absence of electron flow, they are considered to be potential information carriers for low-dissipation, spin-based computing technologies, known as magnonics~\cite{ref:Kajiwara,ref:Magnonics1,ref:Magnonics3,ref:Magnonics5}. The possibility of using waves instead of particles for computing also enables new device concepts for data processing, such as spin wave logic gates~\cite{ref:Gate1,ref:Gate2}. As a first step towards magnonics, it is necessary to encode binary data into spin waves.  Similar to electromagnetic waves, spin waves are characterized by their amplitude, frequency, and polarization. However, in ferromagnets (FMs), the spin wave polarization is always right-handed with respect to the background magnetization. Therefore, one usually has to adopt the spin wave amplitude~\cite{ref:NL} or its frequency to digitize information.

\begin{figure}[!]
	\includegraphics[width=10cm]{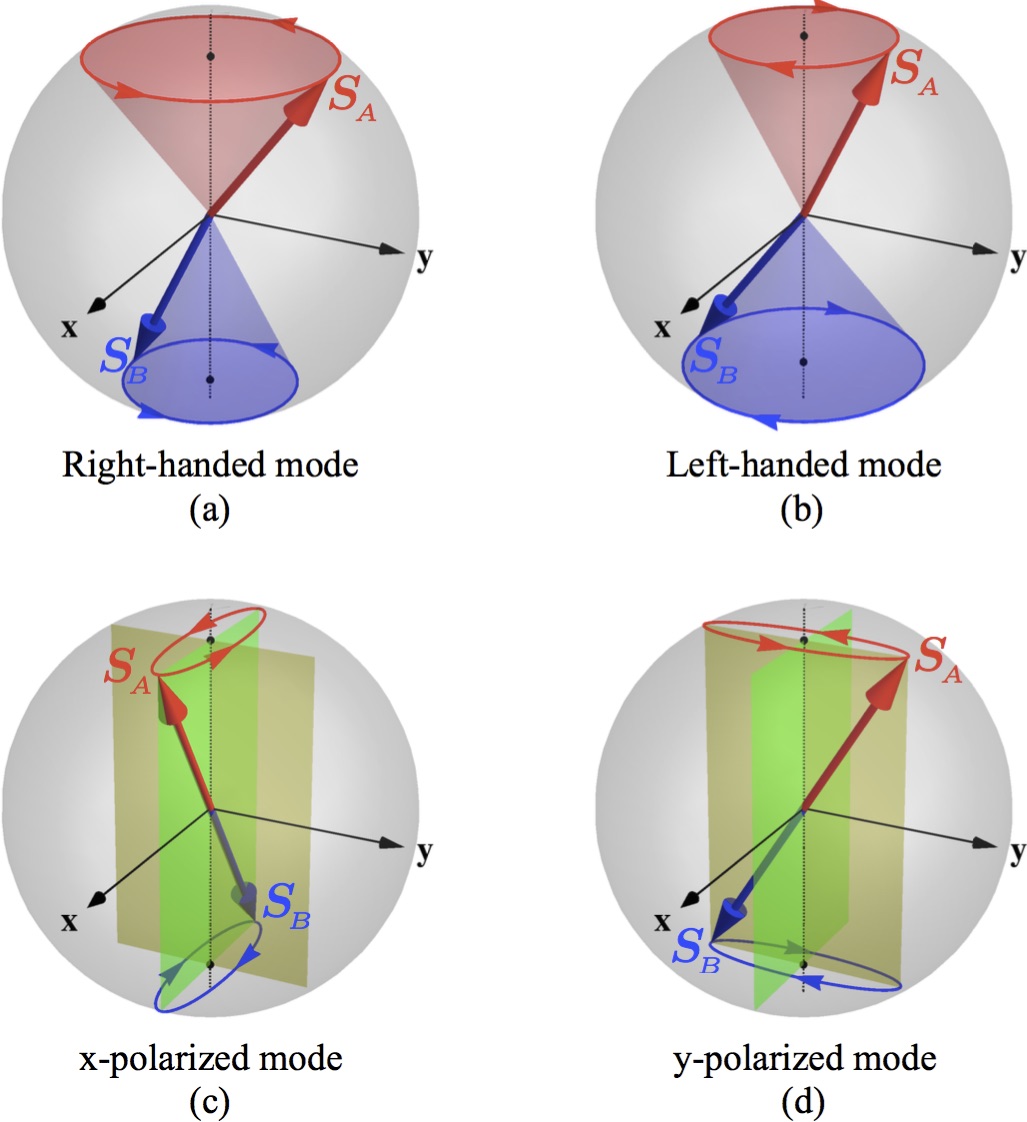}
	\caption{\textbf{Degenerate spin wave modes in a collinear antiferromagnet with easy-axis anisotropy.} Red and blue arrows represent the two sublattice spins $\bm{S}_A$ and $\bm{S}_B$ in a unit cell, with $\hat{z}$ the easy-axis. \textbf{(a)} and \textbf{(b)}: The two circularly polarized modes are characterized by left-handed and right-handed precessions around the easy-axis; they also have slightly different ratios between the cone angles of $\bm{S}_A$ and $\bm{S}_B$. \textbf{(c)} and \textbf{(b)}: The linearly polarized modes consist of different combinations of the circular modes. While $\bm{S}_A$ ($\bm{S}_B$) is individually traveling counterclockwise (clockwise) on an elliptical orbit, the staggered field $\bm{n}=(\bm{S}_A-\bm{S}_B)/2S$ exhibits purely linear oscillation.}
	\label{fig:eigenmode}
\end{figure}

\begin{figure*}
	\includegraphics[width=13cm]{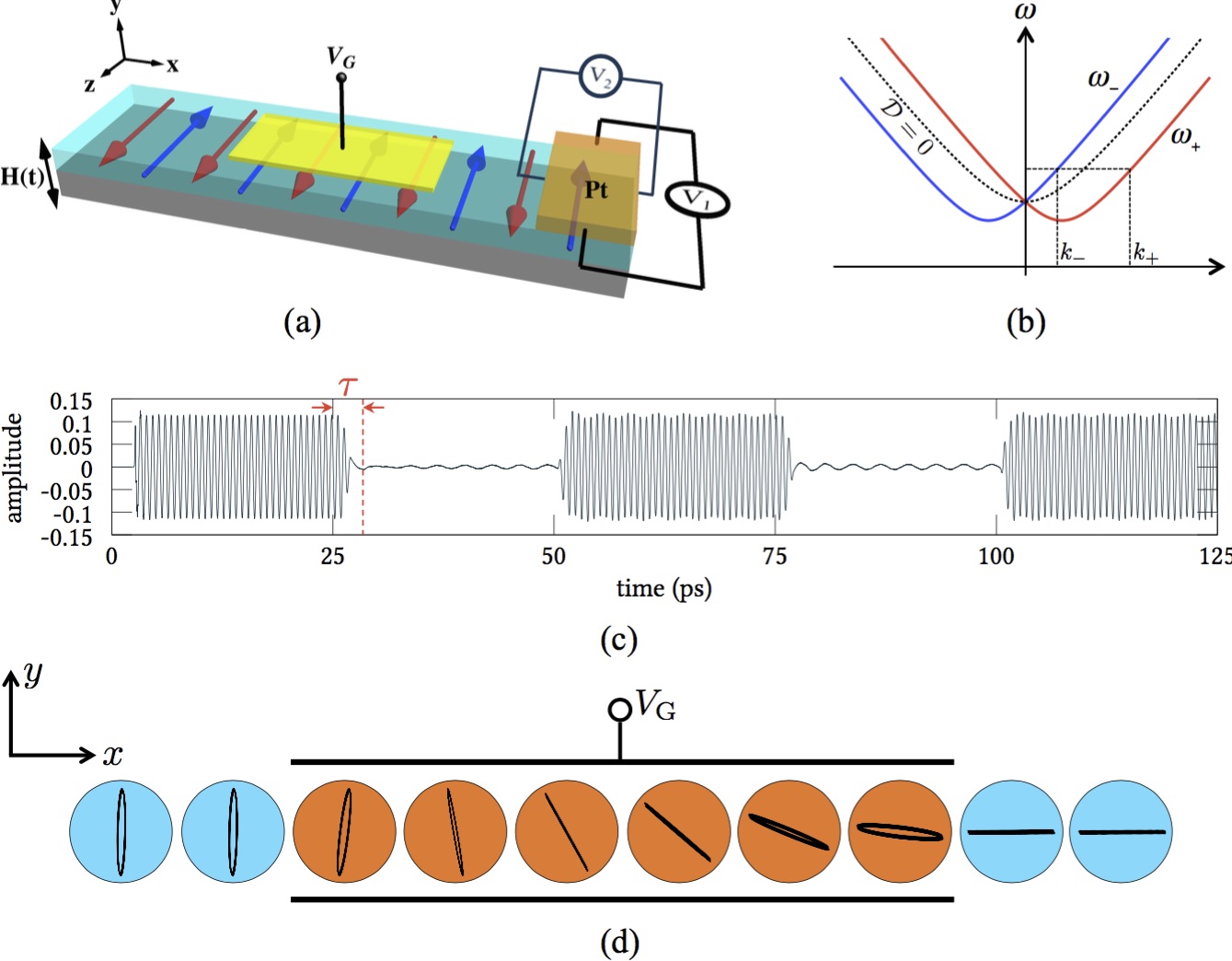}
	\caption{\textbf{A spin wave field-effect transistor and numerical simulation of its performance.} \textbf{(a)} Schematics of the spin wave FET. Spin waves are generated by an oscillating magnetic field $\bm{H}(t)$ at one end of the device and then modulated by a gate voltage $V_G$. At the far end of the device, spin pumping into a heavy metal (\textit{e.g.}, Pt) induces an inverse spin Hall voltage which is measured by two voltmeters $V_1$ and $V_2$. \textbf{(b)} Spin wave spectrum in the presence of the Dzyaloshinskii-Moriya interaction. \textbf{(c)} Numerical simulation of an amplitude shift keying on a AFM nanostrip (see \textbf{Method}). Spin wave signal of 1.4 THz is modulated by a 20GHz square wave from $V_G$. The relaxation time $\tau\approx3.5$ps. \textbf{(d)} Parametric plot of the spatial pattern of the N\'{e}el order for selected sites in and around the gate during the interval $t \in \lbrack40\text{ps}, 50\text{ps})$. }
	\label{fig:device}
\end{figure*}

By contrast, it is well known that the spin wave dynamics of a collinear easy-axis antiferromagnet (AFM) admits two degenerate modes with opposite circular polarization~\cite{ref:AFMR,ref:Keffer}, as illustrated in Figs.~\ref{fig:eigenmode}(a) and (b). These two modes can be recombined into an equivalently orthogonal but linearly polarized basis, as shown in Figs.~\ref{fig:eigenmode}(c) and (d). This two-fold degeneracy places AFM spin waves in a similar situation as electromagnetic waves, which suggests that the spin wave polarization can be harnessed to encode information. In fact, since the magnon chirality is connected to the photon polarization, optical methods have been exploited to excite the two degenerate modes~\cite{ref:Fiebig,ref:Gonokami}. According to recent investigations~\cite{ref:Gomo,ref:AFstt,ref:AFpump}, selective excitation and detection of the two circularly-polarized modes are also achievable via electron spin currents with corresponding polarizations, paving the way to encode information into the polarization of AFM spin waves.

The next crucial step towards AFM magnonics is to control this degree of freedom via external fields in order to perform logical operations on the encoded data. Since the degeneracy is protected by the combined symmetry of time-reversal ($\mathcal T$) and sublattice exchange ($\mathcal I$), a viable control must resort to interactions that break either or both of the two symmetries, i.e., an effective field that couples to the Pauli matrices $\bm\sigma = \{\sigma_1, \sigma_2, \sigma_3\}$ spanning the doubly generate space.

In this article we propose that the Dzyaloshinskii-Moriya interaction (DMI)~\cite{ref:DM1,ref:DM2,ref:Nagaosa} can be used for such a purpose. The DMI is expressed generically as $\bm{\mathcal D}_{AB} \cdot (\bm S_A \times \bm S_B)$, where $\bm{\mathcal D}_{AB}$ is the DMI vector that couples two spins $\bm S_A$ and $\bm S_B$; in an AFM, $\bm S_A$ and $\bm S_B$ represent the two antiparallel moments in a magnetic unit cell. Since the DMI changes sign upon sublattice exchange ($A \leftrightarrow B$), it breaks the degeneracy between the two circular modes. We show that the DMI in AFMs behaves as an fictitious field that couples to $\sigma_3$ in the degenerate space, leading to opposite phase shifts for the two circular modes. As a result, when a linearly-polarized spin wave is subject to the DMI, the opposing phase shifts of its circular components lead to a rotation of the linear polarization direction, which realizes a magnonic analog of the Faraday rotation of electromagnetic waves. If we identify the $x$- and $y$-polarized spin wave modes as 0 and 1 in binary operations, a rotation by $\pi/2$ then corresponds to a NOT operation in magnonic computing.

Based on the Faraday rotation of AFM spin waves, we propose a gate-tunable field-effect transistor serving as the magnonic analog of the Datta-Das device of electrons~\cite{ref:Datta}. We demonstrate its application in the amplitude-shift keying as a THz signal modulation. Finally, by including the field-induced anisotropy~\cite{ref:Banerjee}, we can realize direct transitions between the two circularly-polarized modes, which, together with Faraday's rotation controlled by DMI, enable a complete control of spin wave states over the entire Bloch sphere. Possible electrical detections of the spin wave state on the Bloch sphere are also discussed. Our findings open up the possibility of digital data processing harnessing antiferromagnetic spin waves, and enable the direct projection of optical computing concepts onto the mesoscopic scale. \\

\noindent\textbf{Results}\\
\textbf{Spin wave spectrum.} Under the continuum description, a collinear AFM is characterized by the staggered field $\bm{n}=(\bm{S}_A-\bm{S}_B)/2S$ and the small magnetization $\bm{m}=(\bm{S}_A+\bm{S}_B)/2S$. Consider a quasi-one dimensional nanostrip with an easy-axis along the $\hat{\bm{z}}$-direction, as schematically shown in Fig.~\ref{fig:device}(a). A perpendicular gate voltage is applied to break the inversion symmetry and induce a DMI of the following form (see \textbf{Supplementary S1})
\begin{equation}
H_D=\mathcal{D}[\bm{n}\cdot(\tilde{\nabla}\times\bm{n})+\tilde{\nabla}\cdot(\bm{n}\times\bm{m})-\bm{m}\cdot(\tilde{\nabla}\times\bm{m})], \label{eq:DMI}
\end{equation}
where $\mathcal D$ is the DMI strength and $\tilde{\nabla}=\hat{\bm{y}}\times\nabla$ with $\hat{\bm{y}}$ being the mirror plane normal. The DMI may be nonzero even without gating, as the geometry of the interface already breaks the mirror symmetry. In the exchange limit, $m\ll n$, we will drop the last term; the second term is a total derivative that does not affect the local dynamics. This leaves us with only the first term of Eq.~\eqref{eq:DMI}. The total action in terms of $\bm{n}(x,t)$ is~\cite{ref:NLSM}
\begin{align}
S =\int\mathrm{d}t\mathrm{d}x \, \Bigl[ \mathcal{A}\Bigl(\frac1{c^2}\Bigl|\frac{\partial\bm{n}}{\partial t}\Bigr|^2-\Bigl|\frac{\partial\bm{n}}{\partial x}\Bigr|^2\Bigr)+ \mathcal{K}n_z^2
+\mathcal{D}\Bigl(n_x\frac{\partial n_y}{\partial x}-n_y\frac{\partial n_x}{\partial x}\Bigr)\Bigr], \label{eq:action}
\end{align}
where $\mathcal{A}=Ja/2$ is the stiffness with $J > 0$ the antiferromagnetic exchange coupling and $a$ the lattice constant, $c=2aJ/\hbar$ is the spin wave velocity, and $\mathcal{K}=K/(2a)$ represents the easy axis anisotropy. For small DMI, the (classical) ground state of Eq.~\eqref{eq:action} is the uniform N\'{e}el state. If the DMI exceeds a threshold value $\mathcal{D}_{\text{th}}=2\sqrt{\mathcal{K}\mathcal{A}}=\sqrt{KJ}$, the ground state twists into a spiral. But for our purposes, all discussions are restricted to the sub-threshold regime where the uniform N\'{e}el ground state is preserved.

Linearize Eq.~\eqref{eq:action} in terms of the small deviation $\bm{n}_{\perp}=\{n_x,n_y\}$ of the staggered field from its equilibrium value $\bm{n}^0=\hat{\bm{z}}$, and define $\psi_\pm\equiv n_x\pm in_y$ to associate with the right-handed (left-handed) mode. Setting $\delta S/\delta\psi_{\pm}=0$, we obtain a two-component Klein-Gordon equation
\begin{equation}
\frac{\mathcal A}{c^2}\partial_t^2\Psi = [(\mathcal A\nabla^2 - \mathcal K) + \sigma_3\mathcal D(-i\nabla)]\Psi \;,
\end{equation}
where $\Psi = (\psi_+, \psi_-)^T$.  Solving the above equation with the ansatz $\psi_\pm = \tilde{\psi}_{\pm}e^{i(\omega t-kx)}$ yields the dispersion
\begin{equation}
\omega_{\pm}^2=c^2[k^2\mp Qk+Z], \label{eq:dispersion}
\end{equation}
where $Q\equiv \mathcal{D}/\mathcal{A}$ and $Z\equiv \mathcal{K}/\mathcal{A}$. Equation~\eqref{eq:dispersion} is plotted in Fig.~\ref{fig:device}(b). For an arbitrary given frequency $\omega$ above the antiferromagnetic resonance point at $\omega_R=c\sqrt{Z}$, we find that the splitting of wave vectors
\begin{equation}
\Delta k=k_+-k_-=Q \label{eq:deltak}
\end{equation}
is independent of $\omega$. Let $L$ be the length of the gated region, then the linear polarization of an AFM spin wave will rotate by $\pi/2$ when $\Delta k L=\pi$, which can be regarded as the magnonic analogy of the Faraday rotation of an electromagnetic wave. This mechanism also has a direct analogy in the electron spin field-effect transistor (FET)~\cite{ref:Datta}, where $\Delta k$ is frequency independent and is proportional to the Rashba spin-orbit coupling. \\

\noindent\textbf{AFM spin wave field-effect transistor.} The key to realize an AFM spin wave field-effect transistor is the gate-tunable DMI~\cite{ref:Nagaosa,ref:Vignale1}.  It was shown that in transition metal compounds, an electric field $E$ generates an DMI of the strength $\mathcal{D} = JaeE/\mathcal{E}_{so}$, where $\mathcal{E}_{so}$ originates from the spin-orbit interaction and is typically on the order of 3 eV~\cite{ref:Vignale1}. The electric field $E$ can be produced by either a gate voltage or the interfacial mirror symmetry breaking. Recently, a gate-tunable DMI has been observed in a ferromagnetic insulator~\cite{ref:Zhang}; but the same mechanism is applicable to AFMs as well.  The condition $\Delta k L=\pi$ mentioned above is tantamount to an electric field
\begin{equation}
E=\frac{\pi}{2}\frac{E_{so}}{eL}. \label{eq:Efield}
\end{equation}
If $L\sim1\mu$m, Eq.~\eqref{eq:Efield} is satisfied by $E \sim 3$ V/$\mu$m. The required $E$ field can be scaled down by increasing $L$, but care should be taken that $L$ not exceed the magnon phase coherence length $\ell_\phi$. Though the actual value of $\ell_\phi$ depends on multiple factors, it suffices to consider the magnon dephasing due to the Gilbert damping. This allows for a simple estimate $\ell_\phi\sim v_p/(\alpha\omega)$, where $v_p$ is the phase velocity, $\alpha$ is the Gilbert damping constant, and $\omega$ is the spin wave frequency. Since $v_p=\omega/k$, we have $\ell_\phi\sim1/(\alpha k)$. For $\alpha\sim0.005$ and $k<2\mu$m$^{-1}$, we find that $\ell_\phi$ exceeds 100~$\mu$m. This indicates that spin waves in AFMs can propagate over large distance without losing phase information and provides a loose upper limit for $L$.  On the other hand, the need to suppress spiral formation sets a lower limit for $L$ --- maintaining the stability of the N{\'e}el ground state yields a maximum allowed electric field
\begin{equation}
E_\text{max}=\frac{E_{so}}{ea}\sqrt{\frac{K}{J}}, \label{eq:Emax}
\end{equation}
which depends on the ratio $K/J$ that can differ by several orders of magnitude in different materials ($0.1\sim10^{-4}$)~\cite{ref:MnF2,ref:FeF2,ref:RbMnF3,ref:NaOsO3}. The value of $E_\text{max}$ then sets the corresponding lower bound on $L$ via Eq.~\eqref{eq:Efield}, which can be as short as a few nanometers.\\

\noindent\textbf{THz data modulation.} Since the resonance frequency of AFMs is typically in the THz range, the AFM spin wave FET has important applications in high frequency data modulations. As an example, we perform a numerical simulation of the amplitude shift keying (ASK) based on the device schematic of Fig.~\ref{fig:device}(a). Assume that the spin wave is generated at one end of the chain by an oscillating magnetic field along the $\bm{y}$-direction at 1.4 THz. The spin wave is subsequently modulated by a 20 GHz square wave through the gate voltage $V_G$, which covers 330 atomic sites (see \textbf{Method}). Fig.~\ref{fig:device}(c) plots the $\bm{y}$-component of the transmitted spin wave, where we see a clear on/off ratio.

While the bit rate of such an ASK is as high as THz, the actual signal transfer rate (Baud rate) is limited by the relaxation time $\tau$ of the spin wave FET. As indicated by the dotted red line in Fig.~\ref{fig:device}(b), $\tau$ is the transient period that a transmitted spin wave adapts to an abrupt change of the gate voltage, so the maximum Baud rate cannot exceed $1/\tau$. Physically, the relaxation time can be regarded as the time required for the wavefront of an incident wave to traverse the gated region: $\tau$ must therefore scale as $L/c$. Since $c=2aJ/\hbar$, we have $\tau\sim(L/a)\omega_J^{-1}$ with $\omega_J=J/\hbar$. In typical AFMs, $\omega_J$ is around hundreds of THz, so $\tau$ is only a few picoseconds when $L/a\sim100$. This allows reliable signal transfer at a Baud rate below 0.1 THz. The capacity of signal transfer can in principle be enlarged by reducing $L$ as long as the ground state remains collinear.

Figure~\ref{fig:device}(d) shows the spatial pattern of the staggered field.  Long after the relaxation time, the trajectories of the N\'{e}el order on every 60 atomic sites (30 unit cells) are plotted for several periods. While the transmitted wave is essentially linearly-polarized, the reflected wave and the wave inside the gated region slightly open up into elliptical precessions. This fact implies that the relative ratio between the left-handed wave and the right-handed wave is close to 1 in the transmitted region, whereas it slightly deviates from 1 in the reflected and gated regions. By a straightforward wave matching calculation at the gate boundaries, we find that the deviation of this relative ratio from 1 is linear in $\mathcal{D}/J$ ($=0.5\%$ in our simulation) in the reflected and gated regions, while it is proportional to $(\mathcal{D}/J)^3$ in the transmitted region, which explains the observed pattern (see \textbf{Supplementary S2}).\\



\begin{figure}
	\includegraphics[width=10cm]{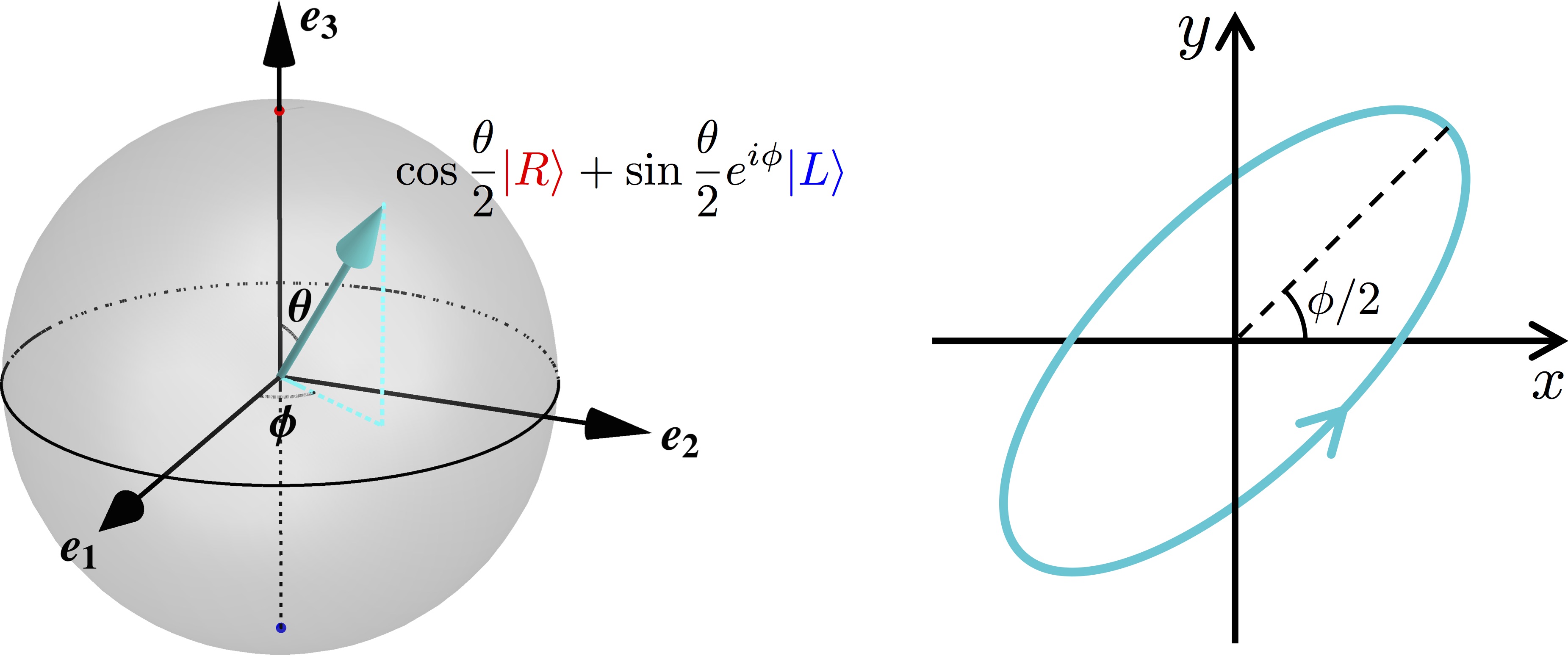}
	\caption{\textbf{Spin wave state on the Bloch sphere}. A spin wave state labeled by ($\theta$, $\phi$) corresponds to an elliptical precession of the N\'{e}el order illustrated in the right panel.}
	\label{fig:digital}
\end{figure}

\noindent\textbf{Spin wave state on the Bloch sphere.} Manipulations of AFM spin waves are not limited to the Faraday rotation, which is realized by the coupling of $\sigma_3$ via the DMI. Coupling to other Pauli matrices extends the attainability of spin wave states to the entire Bloch sphere (see Fig.~\ref{fig:digital}). It has been shown that besides the DMI, an electric field could also induce magnetic anisotropy if it is tilted towards the easy axis~\cite{ref:Banerjee}. This amounts to applying an artificial magnetic field along $\hat{e}_1$ in the internal degenerate space, which rotates the spin wave state in the $\hat{e}_2-\hat{e}_3$ plane, toggling the spin wave chirality. In contrast, the DMI studied above behaves as a magnetic field along $\hat{e}_3$ that rotates the spin wave state along the equator. Under the basis of the circular modes, the equation of motion for spin waves is given by $(\mathcal A/c^2)\partial_t^2\Psi = H\Psi$ with the effective Hamiltonian
\begin{equation}
\begin{split}
H=\Bigl[\mathcal A\nabla^2-\mathcal{K}-\frac{\mathcal{K}'}2\cos^2\vartheta_{_{G}}\Bigr]+\sigma_1\frac{\mathcal{K}'}{2}\cos^2\vartheta_{_{G}} +\sigma_3\mathcal{D}\sin\vartheta_{_{G}}(-i\nabla), \label{eq:Hamiltonian}
\end{split}
\end{equation}
where $\vartheta_{_{G}}$ is the polar angle of the gate voltage direction with respect to the $\hat{\bm{z}}$-axis, and $\mathcal{K}'$ is the strength of the $E$-field induced anisotropy. This Hamiltonian enables us to explore the entire Bloch sphere spanned by the wave function $[\cos\theta/2,\ e^{i\phi}\sin\theta/2]^{\mathrm{T}}$, where $\theta$ and $\phi$ are the spherical angles specifying the position on the Bloch sphere as shown in Fig.~\ref{fig:digital}.

The mapping of the AFM spin wave onto a Bloch sphere places it on an equal footing as the electron spin, in the sense that it can store information through coherent superpositions of two states. Therefore, the AFM spin wave can be regarded as a classical analog of the quantum bit. However, unlike the challenging task of manipulating quantum information, manipulating AFM spin waves is much simpler. For example, by tunning the direction of the gate voltage (i.e., the angle $\vartheta_{_{G}}$ in Eq.~\eqref{eq:Hamiltonian}), the artificial fields that couple to $\sigma_1$ and $\sigma_3$ could be made equal, which mimics the behavior of a Hadamard gate in quantum computing. \\

\noindent\textbf{Detection of spin wave state.} Besides conventional optical approach~\cite{ref:Fiebig,ref:Gonokami}, a spin wave state $(\theta,\phi)$ can also be read off by virtue of spin pumping~\cite{ref:AFpump}. Suppose that a heavy metal is deposited in direct contact to the AFM at the far end of a spin wave FET [see Fig.~\ref{fig:device} (a)]. Spin waves transmitted from the gate can pump spin current into the heavy metal, and this spin current is converted into the inverse spin Hall voltage~\cite{ref:Kajiwara,ref:ISHE} that is monitored by two voltmeters $V_1$ and $V_2$~\cite{ref:Backflow}. By measuring the DC component $\bar V_2$ and the effective AC component $\tilde V_1$ (the root-mean-square value), one can determine $\theta$ and $\phi$ by (see \textbf{Supplementary S3})
\begin{subequations}
\label{eq:detection}
\begin{align}
 \theta&=\arccos\frac{\bar{V}_2}{\bar{V}_2^m}, \\
 \phi&=2\arcsin\notag\left[ \frac{\tilde{V}_1}{\sqrt{2}\tilde{V}_1^m}\frac{\bar{V}_2^m}{\bar{V}_2}\left( \sqrt{1+\frac{\bar{V}_2}{\bar{V}_2^m}}-\sqrt{1-\frac{\bar{V}_2}{\bar{V}_2^m}} \right) \right],
\end{align}
\end{subequations}
where $\bar{V}_2^m$ is the maximum DC voltage along $x$ (for purely right-handed spin wave), and $\tilde{V}_1^m$ is the value of $\tilde{V}_1$ at zero gate voltage (also its maximum).  In principle, the effective AC component can be determined by measuring the output power of the circuit. \\

\noindent\textbf{Discussion}\\
In contrast to FMs, AFMs are devoid of the long range dipolar interaction. In thin film FMs, the dipolar interaction substantially modifies the spectrum in the long wavelength limit~\cite{ref:Stiles}. In particular, the wave vector shift $\Delta k$ due to the DMI tends to zero near the FM resonance frequency~\cite{ref:Vignale2,ref:KD}. In AFMs, however, $\Delta k$ is a constant proportional to the DMI strength as long as $\omega$ is greater than the resonance frequency, as shown above by Eq.~\eqref{eq:deltak}. Therefore, the DMI induces the change of wave vectors more efficiently in AFMs than FMs.

Our proposal for the AFM spin wave FET hinges on the ability to generate sufficiently strong DMI via a gate voltage, which can be realized in compounds with heavy elements.  Another crucial requirement is the availability of suitable AFMs with only easy-axis anisotropy.  Possible candidates include MnF$_2$~\cite{ref:MnF2}, FeF$_2$~\cite{ref:FeF2}, and RbMnF$_3$~\cite{ref:RbMnF3}; all are antiferromagnetic insulators. However, the N\'eel temperatures of these materials are too low (below 100 K) for room temperature applications. A promising replacement is NaOsO$_3$ with the N\'{e}el temperature around 410K~\cite{ref:NaOsO3}. It is a G-type collinear AFM with a dominating easy-axis.  To fully unlock the potential of AFM spin wave FETs thus calls for further development of room-temperature easy-axis AFMs. \\

\noindent\textbf{Methods}\\
The simulations were performed by solving the Landau-Lifshitz-Gilbert equation on a 1D chain using the Dormand-Prince method. We set the saturated staggered magnetization to be unity and scaled everything with frequency. The parameters were then taken as $\omega_J = 100$ THz, $\omega_D = 500$ GHz, and $\omega_K = 10$ GHz. In order to avoid effects on the gate system due to reflection from the chain boundaries, we made the length of the chain much larger than the gate system.

The full chain was constructed with $10^6$ atomic sites ($5 \times 10^5$ AFM unit cells). At the center of the chain, a local magnetic field of $1.4$THz was used as a source of linearly polarized spin waves. Near this source, a 330 site region was gated with DMI. To verify our predictions, we observed the spin wave profiles exiting the far side of the gate. \\

\noindent\textbf{Acknowledgments}\\
We are grateful to Z.-X.\ Gong and J.\ Xiao for helpful discussions.  This study was supported by the U.S. Department of Energy, Office of Basic Energy Sciences, Division of Materials Sciences and Engineering under Award No.~DE-SC0012509 (D.X.) and the National Science Foundation, Office of Emerging Frontiers in Research and Innovation under Award No.~EFRI-1433496 (M.W.D.).\\

\noindent\textbf{Author contributions}\\
D.X.\ and R.C.\ conceived the project. Model calculations were performed by R.C.\ and numerical simulations by M.W.D.\ with input from D.X. and J.Z. R.C.\ and D.X.\ prepared the manuscript. All authors commented on the manuscript.\\

\noindent\textbf{Additional information}\\
\noindent\textbf{Supplementary Information} accompanies this paper at (link available upon publication).

\noindent\textbf{Competing financial interests:} The authors declare no competing financial interests.

\end{document}